\documentclass[12pt,preprint]{aastex}
\usepackage{emulateapj5}
\begin{document}

\newcommand{\kms}{\mbox{km~s$^{-1}$}}
\newcommand{\s}{\mbox{$''$}}
\newcommand{\mloss}{\mbox{$\dot{M}$}}
\newcommand{\mdot}{\mbox{$\dot{M}$}}
\newcommand{\my}{\mbox{$M_{\odot}$~yr$^{-1}$}}
\newcommand{\ls}{\mbox{$L_{\odot}$}}
\newcommand{\um}{\mbox{$\mu$m}}
\newcommand{\ujy}{\mbox{$\mu$Jy}}
\newcommand{\ms}{\mbox{$M_{\odot}$}}

\newcommand{\vexp}{\mbox{$V_{\rm exp}$}}
\newcommand{\vsys}{\mbox{$V_{\rm sys}$}}
\newcommand{\vlsr}{\mbox{$V_{\rm LSR}$}}
\newcommand{\tex}{\mbox{$T_{\rm ex}$}}
\newcommand{\teff}{\mbox{$T_{\rm eff}$}}
\newcommand{\tmb}{\mbox{$T_{\rm mb}$}}
\newcommand{\trot}{\mbox{$T_{\rm rot}$}}
\newcommand{\tkin}{\mbox{$T_{\rm kin}$}}
\newcommand{\dens}{\mbox{$n_{\rm H_2}$}}
\newcommand{\bri}{\mbox{erg\,s$^{-1}$\,cm$^{-2}$\,\AA$^{-1}$\,arcsec$^{-2}$}}
\newcommand{\brib}{\mbox{erg\,s$^{-1}$\,cm$^{-2}$\,arcsec$^{-2}$}}
\newcommand{\flux}{\mbox{erg\,s$^{-1}$\,cm$^{-2}$\,\AA$^{-1}$}}
\newcommand{\ha}{\mbox{H$\alpha$}}

\title{ALMA Observations of the Coldest Place in the Universe: The Boomerang Nebula}
\author{R. Sahai\altaffilmark{1}, W.H.T. Vlemmings\altaffilmark{2},  P. J. Huggins\altaffilmark{3}, L-\AA. Nyman\altaffilmark{4}, 
\& I. Gonidakis\altaffilmark{5}}

\altaffiltext{1}{Jet Propulsion Laboratory, MS\,183-900, California
Institute of Technology, Pasadena, CA 91109, USA}
\altaffiltext{2}{Department of Earth and Space Sciences, Chalmers University of Technology, Onsala Space Observatory, SE-43992 Onsala, Sweden}
\altaffiltext{3}{Physics Department, New York University, 4 Washington Place, New York NY 10003, USA}
\altaffiltext{4}{Joint ALMA Observatory (JAO), Alonso de Cordova 3107, Vitacura, Santiago de Chile, and 
European Southern Observatory, Alonso de Cordova 3107, Vitacura, Santiago, Chile}
\altaffiltext{5}{CSIRO Astronomy and Space Science, Australia Telescope National Facility, Marsfield NSW 2122, Australia}

\email{raghvendra.sahai@jpl.nasa.gov}
\begin{abstract}
The Boomerang Nebula is the coldest known object in the Universe, and an extreme member of the class of Pre-Planetary Nebulae, objects which represent
a short-lived transitional phase between the AGB and Planetary Nebula evolutionary stages. Previous single-dish CO\,(J=1-0) observations (with a 
$45{''}$ beam) showed that the high-speed outflow in this object has cooled to a temperature significantly below the temperature of the cosmic
background radiation. Here we report the first observations of the Boomerang with ALMA in the CO J=2-1 and J=1-0 lines to resolve the structure of
this ultra-cold nebula. We find a central hourglass-shaped nebula surrounded by a patchy, but roughly round, cold high-velocity outflow. We compare
the ALMA data with visible-light images obtained with HST and confirm that the limb-brightened bipolar lobes seen in these data represent hollow cavities
with dense walls of molecular gas and dust producing both the molecular-emission-line and scattered-light structures seen at millimeter and visible
wavelengths. The large diffuse biconical shape of the nebula seen in the visible is likely due to preferential illumination of the cold high-velocity
outflow. We find a compact source of millimeter-wave continuum in the nebular waist -- these data, together with sensitive upper limits on the radio
continuum using observations with ATCA, indicate the presence of a substantial mass of very large (mm-sized) grains in the waist of the nebula.
Another unanticipated result is the detection of CO emission regions beyond the ultracold region which indicate the re-warming of the cold gas, most
likely due to photoelectric grain heating.

\end{abstract}
\section{Introduction}
The Boomerang Nebula, discovered by Wegner \& Glass (1979), holds the distinction of being the coldest known object in the Universe (Sahai
\& Nyman 1997: SN97). The Boomerang is a bipolar Pre-Planetary Nebula (PPN), representing a short-lived ($\sim 1000$ yr) transition
phase during which Asymptotic Giant Branch (AGB) stars and their round circumstellar envelopes (CSEs) evolve into planetary nebulae (PNe)
with a breathtaking variety of aspherical geometrical shapes and symmetries (e.g., Sahai, Morris \& Villar 2011).  The Boomerang's estimated
prodigious mass-loss rate (0.001\,\my) and low-luminosity (300\,\ls) lack an explanation in terms of current paradigms for dusty mass-loss
and standard evolutionary theory of intermediate-mass stars.

Single-dish CO\,(J=1-0) observations (SN97) showed an extended high-speed outflow in absorption against the microwave background, implying
that the nebula has cooled to a temperature
significantly below that of the cosmic background radiation ($T_{cmb}=2.7$\,K) due to adiabatic expansion. Like all PPNs imaged at visible wavelengths at 
high-resolution (e.g., Sahai et al. 2007), the Boomerang appears aspherical, with an hourglass morphology as seen in the light reflected by
dust grains (SN97, also see Hubble Heritage Release 
STScI-2005-25\footnote{http:\//\//hubblesite.org\//newscenter\//archive\//releases\//2005\//25\//image\//b\//}). However, the low angular resolution 
of the CO observations (45\arcsec and 24\arcsec~at J=1-0 and J=2-1, respectively) did not show strong departures from sphericity, 
bringing up
fundamental questions about the relationship between the molecular outflow and the dusty nebula, and thus about its very
formation. High angular resolution mapping of the molecular gas distribution in the Boomerang was needed to resolve the
apparent discrepancy between the molecular and visible-light morphology. This paper reports our mapping of this ultra-cold nebula with ALMA in
Cycle 0 (using the compact configuration) with $\sim2\arcsec-4\arcsec$ resolution, in the CO J=2-1 and J=1-0 lines, 
as well as millimeter-wave continuum emission at 1.3 and 2.6\,mm. 
We compare these high-resolution images of the molecular gas with archival visible-light images obtained with the Hubble Space Telescope (HST). We also
report radio continuum observations with ATCA that we obtained in order to help our understanding of the millimeter-wave continuum.

The plan of the paper is as follows. In \S\,\ref{obs} we describe the observational setups, and data reduction and calibration procedures. In 
\S\,\ref{result} we present our main observational results, as derived from the visible-light data (\S\,\ref{opt}), the CO\,(J=2-1) and CO\,(J=1-0) line
data
(\S\,\ref{coj21} and \ref{coj21}), and the continuum data (\S\,\ref{contxt}). In \S\,\ref{discus}, we discuss and analyse these results, and in 
\S\,\ref{conclude} we present the main conclusions of our study. We adopt a value of 1.5\,kpc for the distance to the Boomerang (as inferred by SN97).

\section{Observations}\label{obs}
The CO\,(J=1-0) emission of the Boomerang Nebula, at a rest frequency of $115.2712$~GHz, and the nearby continuum, were observed using ALMA band
3 on November 29
and December 31, 2011. The CO\,(J=2-1) emission line at $230.3581$~GHz, and the nearby continuum, was observed, using ALMA band 6, on November 28
2011. In both
cases, the data contains four spectral windows of $1.875$~GHz width and with 3840 channels. The band 3 observations have spectral windows
centered on approximately $100.7, 102.6, 112.7$, and $114.6$~GHz while the band 6 observations have windows centered on $228.4, 230.4,
242.4$ and $244.4$~GHz. The channels spacing of $0.488$~MHz corresponds to $1.27$~km~s$^{-1}$ for the CO\,(J=1-0) and $0.63$~km~s$^{-1}$ for
the CO\,(J=2-1) line. The data was taken using the compact ALMA cycle 0 configuration, with baselines ranging from 14~m up to 200~m.

In the case of the CO\,(J=1-0) observations, a 7-point mosaic, using 15 antennas in November and 18 antennas in December, was used to map a
circular area with a radius of $\sim0.9'$ centered on R.A.$=12^h44^m45.449^s$ and Dec$=-54^\circ31'11.388{''}$. The CO\,(J=2-1) was observed
with a single pointing at the central position using 15 antennas of which two were flagged due to problems with the water vapour radiometer
(WVR). The pointing position corresponds to the Simbad coordinates of the Boomerang Nebula,  R.A.$=12^h44^m45.45^s$
and Dec$=-54^\circ31'11.4{''}$.
The total observing time for the CO\,(J=1-0) line was 3.25~hours with each mosaic pointing observed for
$\sim19$~min in total. The single pointing of the CO\,(J=2-1) line was observed for only $6$~minutes. In both cases, bandpass and gain
calibration were performed on the quasar J1329-5608, while flux calibration was done using Mars. The flux calibration was bootstrapped
between spectral windows affected by line emission from Mars itself to the gain calibrator. We measured a flux of
$1.63\pm0.02$~Jy~beam$^{-1}$ for J1329-5608 in November in band 3, which increased to $1.80\pm0.03$~Jy~beam$^{-1}$ in December. In band 6,
we measured a flux of $1.04\pm0.04$~Jy~beam$^{-1}$. We estimate the final absolute flux calibration to be accurate to within $3\%$ in band
3 and $5\%$ in band 6.

The data were reduced using the Common Astronomy Software Application (CASA 3.4.0). After corrections for time and frequency dependence of
the system temperatures and rapid atmospheric variations at each antenna using WVR data, we improved the antenna positions. Subsequently,
bandpass and gain calibration were done, and the calibration solutions determined on J1329-5608 were applied to the Boomerang Nebula.
Imaging was done using the CASA clean algorithm after a continuum subtraction was performed on the emission line data. As a significant
amount of flux of the CO\,(J=1-0) line was resolved out, we performed various cleaning rounds using different weighing schemes (from natural
to uniform) and different data tapering to assess the robustness of the observed structure. As consistent structure was recovered in the
different cleaning runs, we conclude that these are not artifacts of the missing short spacings. Our final CO\,(J=1-0) image was created
by channel averaging to a spectral resolution of $6.25$~km~s$^{-1}$, using natural weighing, a Gaussian taper of $50$~k$\lambda$ and a
restoring beam of $4\farcs5\times3\farcs0$ at a position angle of $-30^\circ$. In the line-free channels, the resulting rms noise was 
$\sim2.5$~mJy~beam$^{-1}$. The CO\,(J=2-1) map was imaged at a spectral resolution of $0.63$~km~s$^{-1}$, Briggs weighing and a Gaussian
taper of $100$~k$\lambda$; the beam was $2\farcs40\times1\farcs55$ at a position angle of $-3.5^\circ$. In the line-free channels, this resulted in an
rms noise of $\sim7.5$~mJy~beam$^{-1}$. Finally, we combined a
CO\,(J=1-0) image cube with a spectral resolution of $8$~km~s$^{-1}$ with the original SEST observations (SN97) using the CASA task {\it
feather}, in which the regridded image data is combined in the Fourier plane. A lower spectral resolution was used to reduce the noise
contribution from the SEST observations.

HST images, taken with the ACS/HRC and WFPC2 instruments, through the F606W wide-band filter, on March 25, 2005, and March 23, 1998 via programs
GO\,10378 and GO\,6856, were extracted from the MAST archive and the Hubble Legacy Archive, respectively. For the HRC data two 160\,sec exposures
were registered and combined in order to remove cosmic-rays. For the WFPC2 data, the Legacy Archive provided an average image combining two 500\,sec
exposures. Since part of the extended nebula is obscured by the occulting finger of the ACS/HRC coronagraph, we used the WFPC2 image (in which the
central region is saturated) for comparing the large scale visible-light morphology with our ALMA data. 

We observed the Boomerang Nebula in Director's Discretionary Time (DDT) with the Australia Telescope Compact Array (ATCA) on June 21, 2012. A 10-h
observation run (project code CX241) was carried out with the 6D array configuration, using 2\,GHz bandwidths at 5.5, 9, 22, and 24\,GHz with the
compact array broadband backend (CABB). The calibration and the imaging of the data were done using the MIRIAD package. Bandpass and flux calibration
were performed on the standard calibrator 1934-638 and phase calibration was performed on 1326-5256. The fluxes are in agreement with the values
presented in the ATCA calibrators database, within an uncertainty of less than $10\%$. After calibration, the imaging of the source was performed
using multifrequency synthesis with a single continuum image produced for the 22\,and 24\,GHz bands (identified with the central frequency of
23\,GHz). The beam size of the observations was $2\farcs5\times3\farcs1$, $1\farcs5\times1\farcs9$, and $0\farcs6\times0\farcs7$ for the 5.5, 9, and
23\,GHz observations respectively.

\section{Results}\label{result}
\subsection{Visible-light imaging}\label{opt}
\begin{figure}[!ht]
\resizebox{1.0\textwidth}{!}{\includegraphics{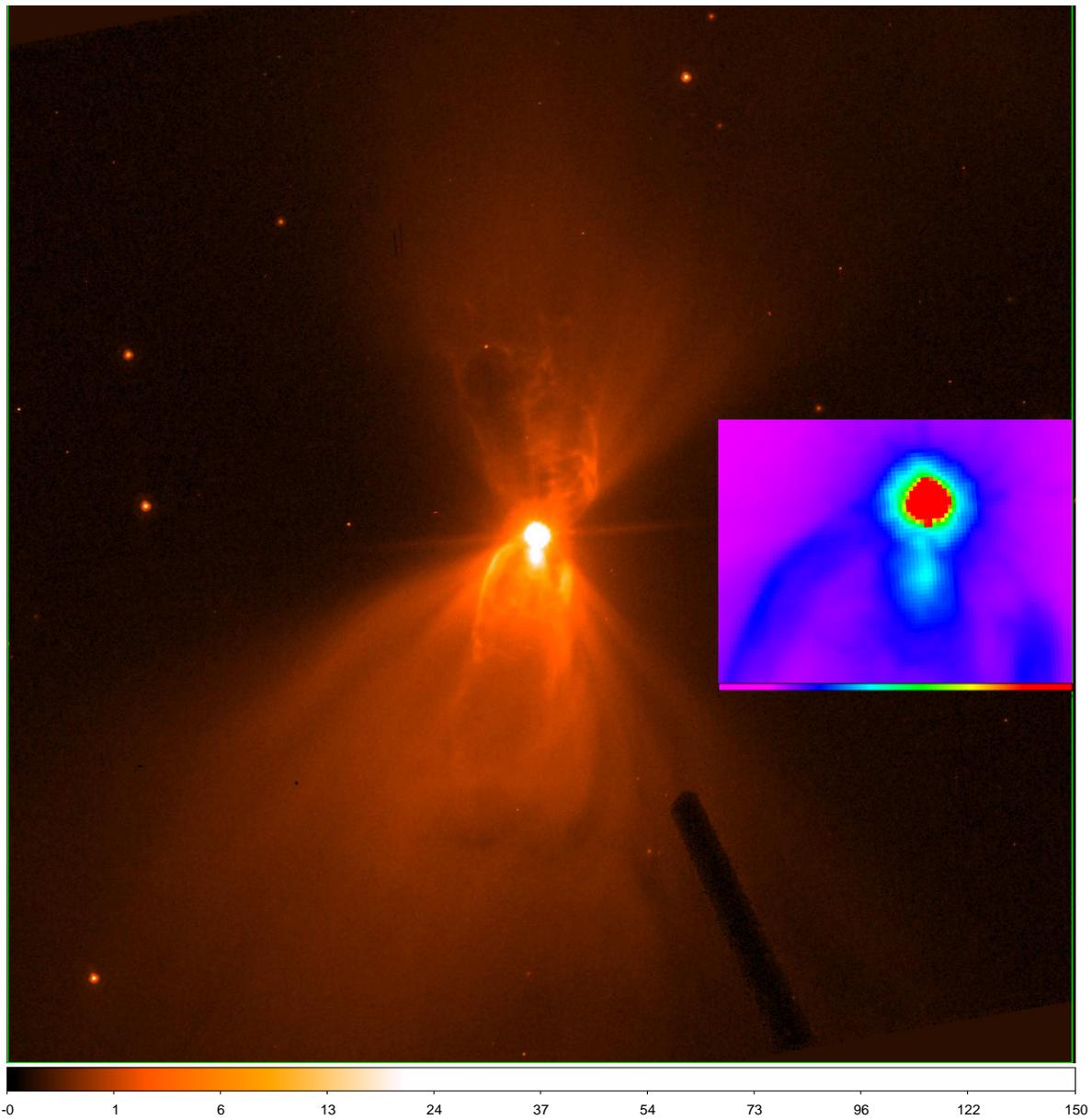}}
\caption{HST F606W image ($24{''}\times24{''}$, shown using a square-root stretch. Insert is a magnified view of the central region, showing the 
presence of two knots along the central linear feature in the S-lobe. The dark linear feature at lower right is due to the occulting finger of the 
ACS/HRC coronagraph.}
\label{hst606w}
\end{figure}

We describe the HST imaging first, as this shows the structural details of the nebula (in scattered light), at the highest angular resolution
($\sim$0\farcs05; over-sampled by a factor 2 in the HRC with a plate scale of 0\farcs025/pixel.)
The HRC image (Fig.\,\ref{hst606w}) shows a limb-brightened lobe in the south, that, at its base, flares out from the center in a
wide-V shape out to an axial distance of $0\farcs7$, and then follows a roughly cylindrical shape, i.e., the western and eastern limbs are
parallel, out to about an axial offset of $4\farcs4$, beyond which they curve inwards and merge; the resulting lobe-end  
is located at an axial offset of $5\farcs6$. The lobe axis is oriented at $PA\sim173\arcdeg$. 

The northern lobe, also limb-brightened, is much fainter and has a more complex structure. There is a wide V-shaped flared structure as
in the south, that then transitions towards a more cylindrical shape, however the western and eastern limbs of this lobe are not exactly parallel. The
average PA of the long axis of the N-lobe is about $6\arcdeg$, and thus not aligned with that of the S-lobe. The lateral width of the N-lobe
(about $2\farcs5$, measured at an axial offset of $2{''}$) is larger than that of the 
S-lobe ($2\farcs1$).  The W- and E- limbs of the N-lobe curve inwards towards the lobe-axis and merge, and the end of the lobe is located 
at an axial offset of $4\farcs2$. 

Each limb-brightened lobe appears embedded in a
larger, more diffuse conical nebulosity with an opening angle that is the same as that of its wide V-shaped inner region 
(measuring about 90\arcdeg~in the north and 100\arcdeg~in the south). 

A bright linear feature can be seen in the middle of the S-lobe, emanating from the central star, extending out to an offset of about $1{''}$, at
$PA=180\arcdeg$. This feature has significant
structure, the most noteworthy being the presence of two bright knots, one at a radial distance, $r=0\farcs43$, $PA=179\arcdeg$ from the center, and
another slightly fainter and smaller, located at $r=0\farcs23$, $PA=172\arcdeg$. 

\subsection{CO\,(J=2-1)}\label{coj21}
\begin{figure}[!ht]
\vskip -1.7in
\resizebox{1.1\textwidth}{!}{\includegraphics{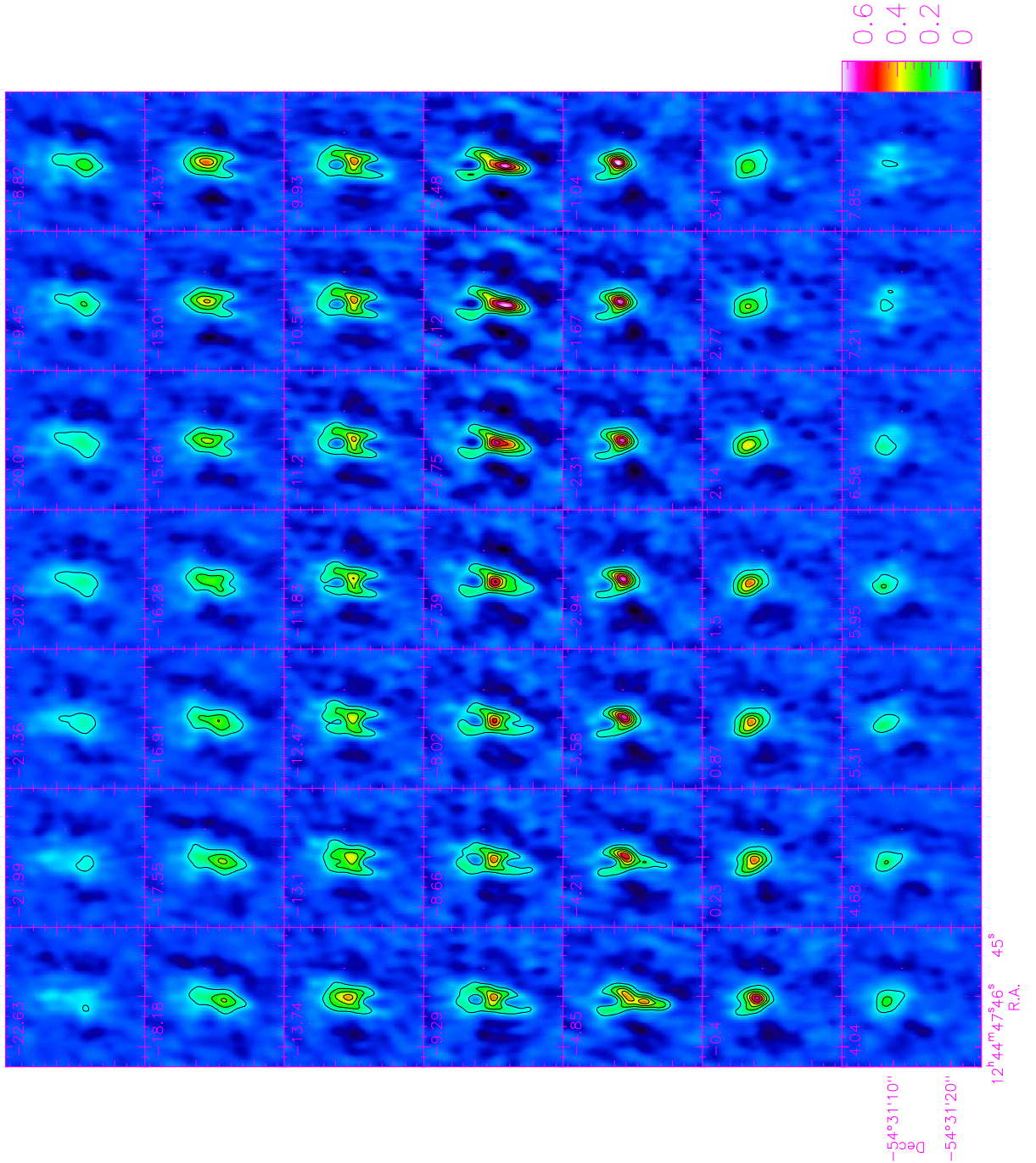}}
\vskip -0.8in
\caption{ALMA CO\,(J=2-1) channel maps of the Boomerang, covering the radial velocity range $V_{lsr}=-22.63$ to $7.85$\,\kms. Contours levels 
are 0.1, 0.2, 0.3, 0.4, 0.5, and 0.6 Jy/beam, and the color-coding of the intensity scale (in Jy/beam)
is shown at the bottom right corner. The central velocity for each panel is shown in its top-left corner, and the channel width is 0.63\,\kms. }
\label{co21chan}
\end{figure}

\begin{figure}[!ht]
\vskip -3in
\resizebox{1.1\textwidth}{!}{\includegraphics{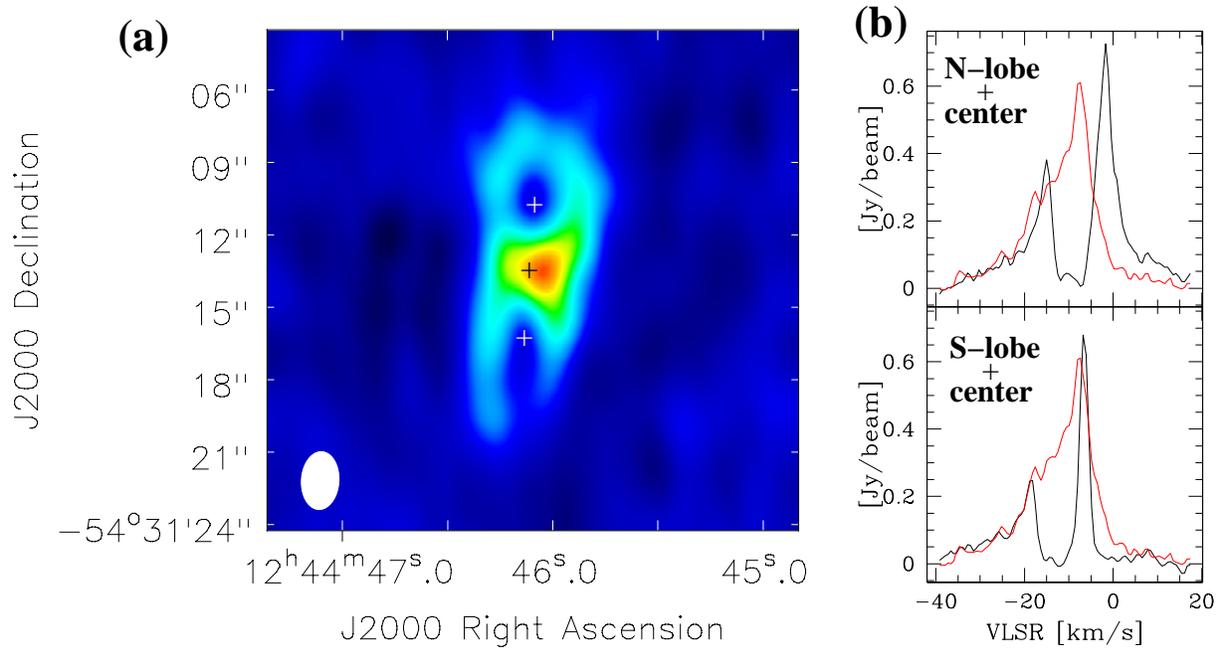}}
\caption{ALMA CO\,(J=2-1) map (at the systemic velocity) ({\it panel a}), and representative spectra ({\it panel b}) of the Boomerang. The latter have
been  
extracted from the waist center ({\it black cross}) and the northern and southern lobes ({\it white crosses}). The waist spectrum is overlaid 
(in red) on each of the lobe spectra (in black) for comparison.
%
}
\label{co21map}
\end{figure}

We show the CO\,(J=2-1) emission as a function of radial velocity in Fig.\,\ref{co21chan}.
At the systemic velocity ($\sim-10$\,\kms: SN97), the CO\,(J=2-1) image represents a cut of the density structure in the sky plane, and shows a
roughly hourglass-shaped, bipolar nebula in emission with a central waist. The 
two limb-brightened lobes defining the hourglass cover a region of about $13\farcs6\times\,4\farcs4$
(see Fig.\,\ref{co21map}a); the waist has a lateral extent of about $3\farcs8$ (as measured from an intensity cut across the waist between locations
where
the intensity drops to half its value at the geometrical center of the waist). Spectra extracted from representative locations within the lobes show
double-peaked profiles (Fig.\,\ref{co21map}b), as expected due to emission from the front and back parts of lobes that are hollow in their 
interior.  The mean
velocity for the S-lobe is blue-shifted from that of the N-lobe, implying that the southern (northern) lobe is tilted towards (away from) us.  In
contrast, the profile towards the center of the nebula, is centrally-peaked. The width of the profiles at their base (FWZI) is about 65\,\kms.

We do not expect that there is any loss of flux in our CO\,(J=2-1) image of the Boomerang because the total emission extent is smaller than the
angular
scale at which the ALMA observations are expected to resolve out structures ($\gtrsim17{''}$), and consistent with this expectation, 
we find that the spatially-integrated ALMA CO\,(J=2-1) flux ($3$\,Jy) is not significantly different from the single-dish (SEST) flux ($3.1$\,Jy). 

\begin{figure}[!ht]
\resizebox{1.0\textwidth}{!}{\includegraphics{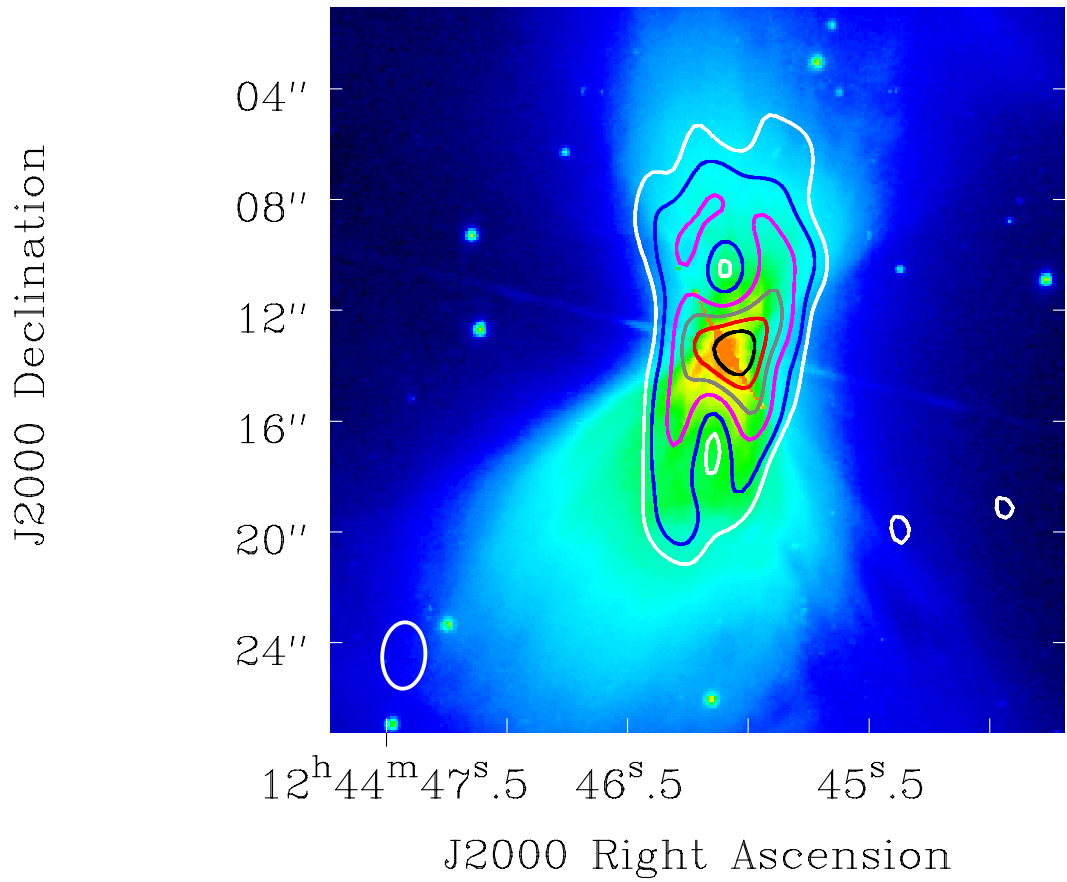}}
\caption{ALMA CO\,(J=2-1) map (contours) of the Boomerang at the systemic velocity overlaid on false-color HST/WFPC2 image taken
with filter F606W. Contours levels shown are at 5\,$\sigma$ (white), 10\,$\sigma$ (blue), 20\,$\sigma$ (magenta), 30\,$\sigma$ (grey), 40\,$\sigma$
(grey), and 50\,$\sigma$ (black), with $\sigma=7.5$\,mJy/beam. The beam (FWHM) is shown as the white ellipse in the lower left corner.
}
\label{hstco21}
\end{figure}

The long axis of the nebula seen in CO  
is roughly aligned with the symmetry axis of the nebula seen in the WFPC2 image (Fig.\,\ref{hstco21}). 
We find detailed association between features in the CO\,(J=2-1) image and the HST image. Like the latter, the CO\,(J=2-1) map at the systemic
velocity
shows an hourglass structure. The E- and W- peripheries of this structure appear roughly cylindrical with an orientation that is
consistent with that of the visible-light 
S-lobe, in several channels at and near the systemic velocity. However, at larger red-shifts relative to the systemic velocity, i.e., for
$V_{lsr}\gtrsim-5$\,\kms, the PA appears to shift 
anti-clockwise, bringing it into rough alignment with the W-limb of the visible-light N-lobe. In summary, in spite of the large difference in the resolution 
of the visible-light and mm-wave images that precludes a more precise comparison between the two, the ALMA data are consistent with the 
bulk of the CO emission arising from the walls of the lobe cavities seen via scattered-light in the HST image.

\subsubsection{Spatio-Kinematic Structure of the Lobes}
\begin{figure}[!ht]
\rotatebox{270}{\resizebox{0.6\textwidth}{!}{\includegraphics{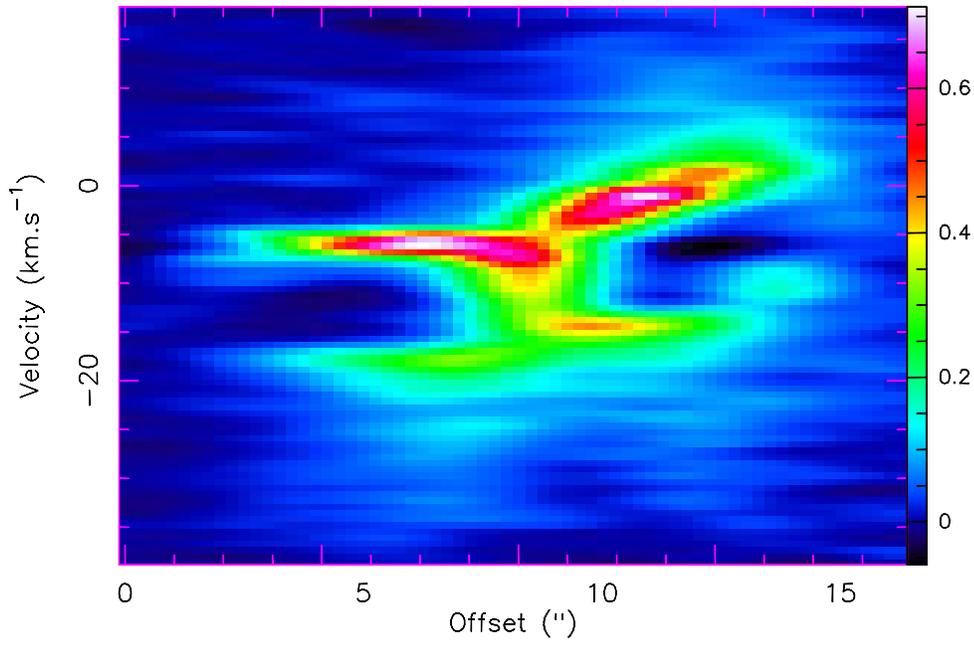}}}
\caption{CO\,(J=2-1) position-velocity diagram of intensity along the major axis of the nebula. Offsets are measured 
from south to north, beginning at R.A.$=12^h44^m46.21^s$, Dec$=-54^\circ31'22.1{''}$ and terminating at R.A.$=12^h44^m46.02^s$, 
Dec$=-54^\circ31'02.4{''}$.
The color-coding of the intensity scale (in Jy/beam) is shown on the right side of the panel.}
\label{co21major}
\end{figure}

The position-velocity plot of the CO\,(J=2-1) emission for a cut along the major axis of the nebula (Fig.\,\ref{co21major}) also reveals an
hourglass-shaped structure, and shows the emission from the front wall (red-shifted) and back wall (blue-shifted) of the north and south lobes. The
emission from each wall of the S-lobe is blue-shifted relative to the corresponding wall of the northern lobe, confirming our inference above 
that the southern
(northern) lobe is tilted towards (away from) us.  The tilt of the lobes inferred above is consistent
with the larger overall visible-light brightness of the southern lobe compared to the northern one -- we expect the lobe closer to us to be  
brighter because for it, (a) the scattered light traverses a
smaller column density of foreground circumstellar dust, (b) the starlight is scattered
more efficiently towards the line-of-sight if the grains have a forward-peaked scattering phase function. 

Since all emission from the S-lobe's far wall is red-shifted from the systemic velocity, we can set an upper limit of $15\arcdeg$ on the inclination
of the nebula axis to the sky-plane, assuming radial expansion and a cylindrical geometry for the S-lobe, using the S-lobe's diameter (3\farcs3) (as
measured from the
lateral separation of the mid-point of each wall in the CO\,(J=2-1) image at the systemic velocity) and maximum projected length (6\farcs5) (as
measured for the red-shifted emission feature in the S-lobe in Fig.\,\ref{co21major}).

For similar velocity offsets from the systemic velocity, the red-shifted emission is, in general, significantly brighter than the corresponding
blue-shifted one (e.g., Fig.\,\ref{co21major}). This effect is most simply understood if (a) there is a temperature gradient across the thickness of
the lobe walls, such that the interior surface of the walls is hotter than the exterior one, and (b) the CO\,(J=2-1) emission is optically-thick.  
For optically-thick emission at any given velocity offset, we can only see emission from the surface of each wall that is
closest to us: hence, at red-shifted (blue-shifted) velocities, we see the hotter, interior (cooler, exterior) surfaces of the lobe walls. 

We expect the lateral separation between the walls of each lobe in the channel maps to progressively decrease with increasing offset from the systemic
velocity, until they finally merge along the nebular axis (Fig.\,\ref{co21chan}). This ``narrowing" effect is seen most clearly for the S-lobe, where
a prominent narrow
linear feature appears at $PA\sim175\arcdeg$ in the velocity range $V_{lsr}=-7.39$\,\kms~to $-4.21$\,\kms, with a linear extent of about $5\farcs9$ as
measured at the half-intensity points at $V_{lsr}=-5.48$\,\kms. The blue-shifted equivalent of this feature for the S-lobe, is seen in the velocity
range 
$V_{lsr}=-16.91$\,\kms~to $-18.18$\,\kms, but is fainter (since it comes from the exterior, cooler surface of the lobe) and less extended. The same
``narrowing" effect can also be seen for the blue-shifted emission from the N-lobe (e.g., at $V_{lsr}=-16.28$\,\kms~to $-19.45$\,\kms). But
the ``narrowing" effect and the equivalent red-shifted feature is not seen for the far wall of the N-lobe. The spatio-kinematic structure of the
N-lobe is clearly more complex than that of the S-lobe.


\subsubsection{Spatio-Kinematic Structure of the Waist}
\begin{figure}[!ht]
\rotatebox{270}{\resizebox{0.6\textwidth}{!}{\includegraphics{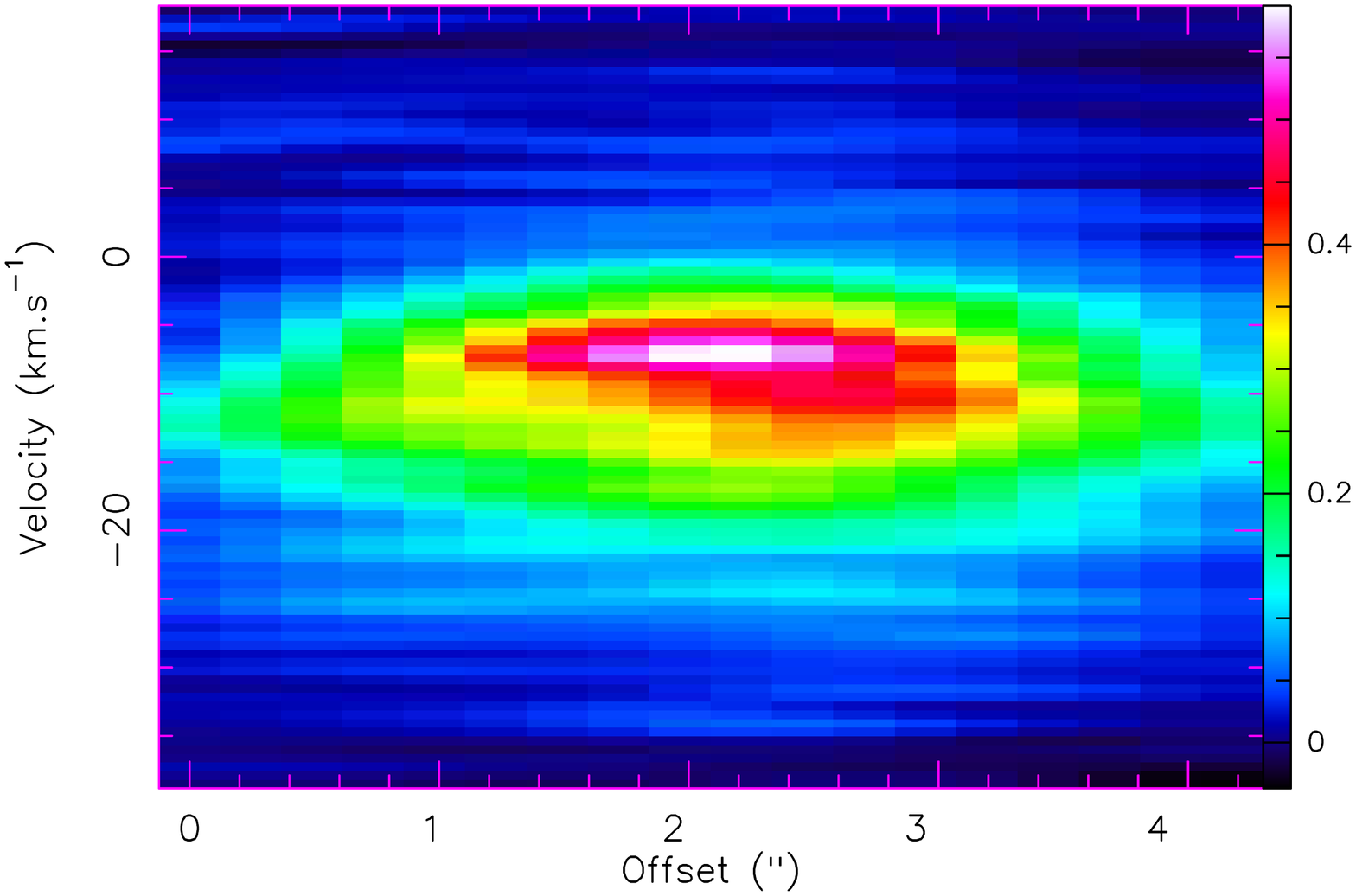}}}
\caption{CO\,(J=2-1) position-velocity diagram of intensity along a cut traversing the waist of the Boomerang. Offsets are measured 
from east to west, beginning at R.A.$=12^h44^m46.31^s$, Dec$=-54^\circ31'12.8{''}$ and terminating at R.A.$=12^h44^m45.81^s$, 
Dec$=-54^\circ31'13.4{''}$.
The color-coding of the intensity scale (in Jy/beam) is shown on the right side of the panel.}
\label{waistpv21}
\end{figure}

In a position-velocity (PV) plot of a cut across the waist in the CO\,(J=2-1) map (Fig.\,\ref{waistpv21}), one can see evidence for a
velocity-gradient
in the line centroid, which appears to shifts steadily, from about $-7$\,\kms~at an offset of 
about $\sim1\farcs2$ to $-11$\,\kms~at an offset of $\sim2\farcs8$. In contrast, at low intensity levels, the emission is symmetric
(green region in plot). The gradient appears to be real and not an artifact of the limited resolution mixing in emission from the N-S outflow,
as this would result 
in an even more pronounced gradient in the green region. But the interpretation of the PV plot (e.g., as resulting from rotation) is not
straightforward -- e.g. the symmetric
green outer region has its major axis at $-12$\,\kms, but the mid-point of the asymmetric orange-red region in the center appears at
$\sim-8$\,\kms.

\subsection{CO\,(J=1-0)}\label{coj10}
\begin{figure}[!ht]
\rotatebox{90}{\resizebox{1.3\textwidth}{!}{\includegraphics{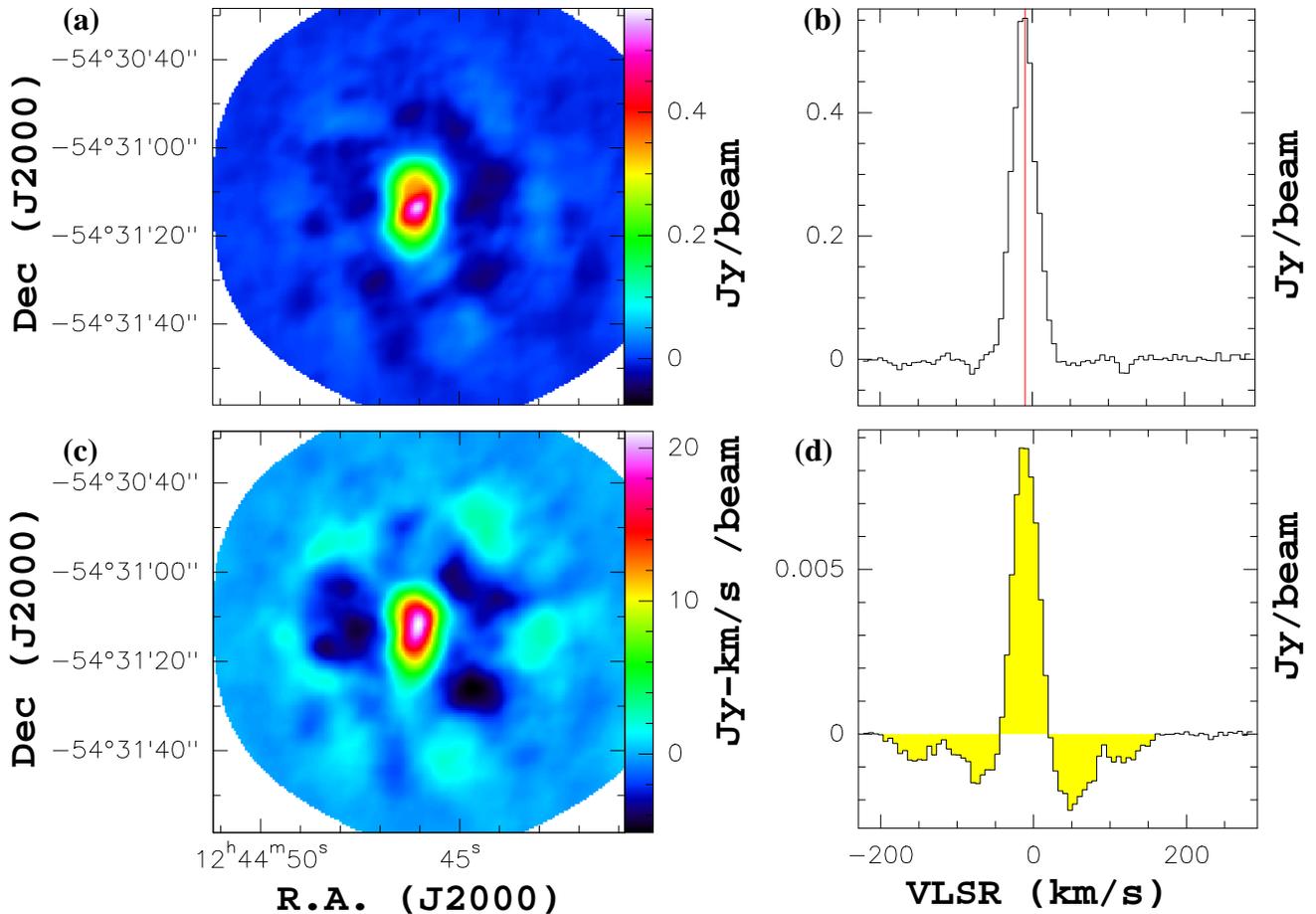}}}
\caption{ALMA CO\,(J=1-0) map and spectra of the Boomerang: $(a)$ map of the
Boomerang at the systemic velocity; $(b)$ spectrum extracted from the waist-center;
$(c)$ map integrated over the velocity range, $V_{lsr}=-200$ to $180$\,\kms~(shown in {\it yellow}); $(d)$, spectrum averaged over the field-of-view
in panel $c$ (the
velocity range for the map in panel $c$ is shown in yellow). 
}
\label{co10map}
\end{figure}

The J=1-0 image (Fig.\,\ref{co10map}) shows a central bipolar nebula in emission, roughly similar
to that seen in the J=2-1 line; but due to the factor 2 lower resolution at 2.6\,mm, the limb-brightened structure is not resolved. The
bipolar nebula is surrounded on all sides by large patchy regions of absorption. The absorption
regions lie roughly within a circle of diameter $\sim50{''}$. Immediately beyond the absorption region, one can
see faint patchy emission regions, lying within a circle of diameter $\sim65{''}$. The spectrum towards the center of the nebula 
(Fig.\,\ref{co10map}b) shows a strong 
emission peak centered at the systemic velocity with some weak absorption features, due to regions in the ultra-cold outflow that lie along the
line-of-sight to the center. The FWZI of the emission component is about 85\,\kms, somewhat larger than that of the central emission component
measured in the CO\,(J=2-1) line. In contrast, the spectrum (Fig.\,\ref{co10map}d) averaged over the whole nebula shows, in addition to the central
emission, absorption features extending to expansion velocities up to 170 (190)\,\kms~redwards (bluewards) of the systemic velocity.

\begin{figure}[!ht]
\resizebox{0.6\textwidth}{!}{\includegraphics{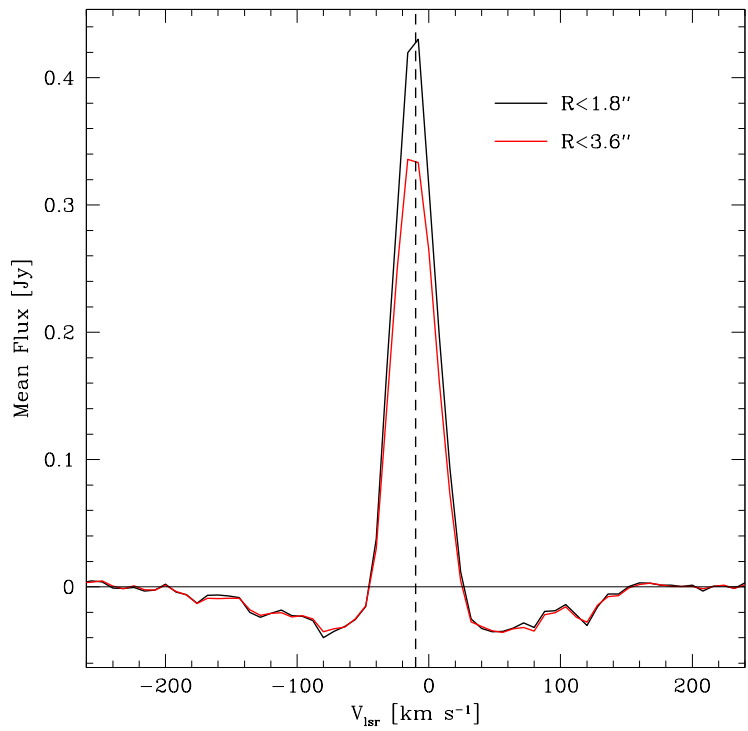}}
\caption{The ``line-of-sight" CO\,(J=1-0) spectra towards the center of the Boomerang, extracted from the ALMA+SEST map. The spectra have been
extracted from (and averaged over) two circular apertures, one with diameter equal to the mean beam FWHM, and the other twice the mean beam FWHM.
}
\label{almasest_spec}
\end{figure}

The CO\,(J=1-0) spectra towards the center of the nebula, extracted from the ALMA+SEST map (Fig.\,\ref{almasest_spec}), shows absorption over a wide
range of velocities, different from the expectation of a shell expanding at a constant velocity, in which case the spectra would show absorption only
at and near the outflow velocity. The spectra have been extracted from (and averaged over) two circular apertures, one with diameter equal to the
mean beam FWHM, and the other twice the mean beam FWHM. If the absorption at relatively low outflow velocities was simply because of a projection
effect, i.e., due to high-velocity material close to the center that is included within the aperture but is expanding at a large angle to the
line-of-sight, then we would expect the intensity of the absorption feature in the larger-aperture spectrum to be significantly stronger than in the 
smaller-aperture one, contrary to what is observed\footnote{the average emission feature is weaker for the larger-aperture because the central
hourglass nebula becomes generally fainter with distance from the center}.

\begin{figure}[!ht]
\resizebox{0.6\textwidth}{!}{\includegraphics{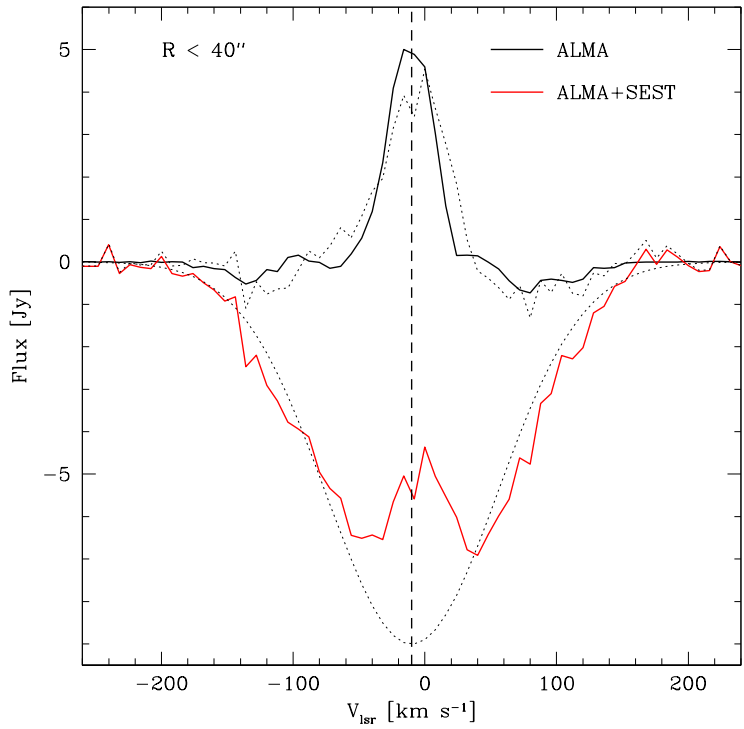}}
\caption{Spatially integrated CO\,(J=1-0) spectrum of the Boomerang over a circular aperture of radius $40{''}$: {\it (top)} ALMA data
only and {\it (bottom)} ALMA+SEST data. The dotted absorption curve is a
gaussian fit to the wings of the absorption feature, and the dotted emission
curve is the difference between the ALMA+SEST spectrum and the former.
}
\label{bothspec}
\end{figure}

We expect a large fraction of the CO\,(J=1-0) absorbing cloud to be 
resolved out in our ALMA data, because in the configuration used for our observations, the array was not sensitive to
structure on scales $\gtrsim35{''}$. As a result, we find the absorption feature in the CO\,(J=1-0) spectrum derived 
from spatially-integrating over the full source, is
much weaker in the ALMA map compared to the corresponding spectrum from the ALMA+SEST map (Fig.\,\ref{bothspec}).

However, we do not expect the central bipolar source seen in emission to be over-resolved, and so the signal from this 
component should have the same intensity in both the ALMA and ALMA+SEST maps. This expectation is 
supported by our finding that the peak flux of the emission component in the 
ALMA spectrum (5\,Jy) is only about 20\% different from (larger than) that in the ALMA+SEST spectrum (as measured above a gaussian fit to the
underlying shape of the absorption component), a discrepancy that is well within the calibration uncertainties of both data sets.




\subsection{Continuum}\label{contxt}
\begin{figure}[!ht]
\vskip -3in
\resizebox{1.1\textwidth}{!}{\includegraphics{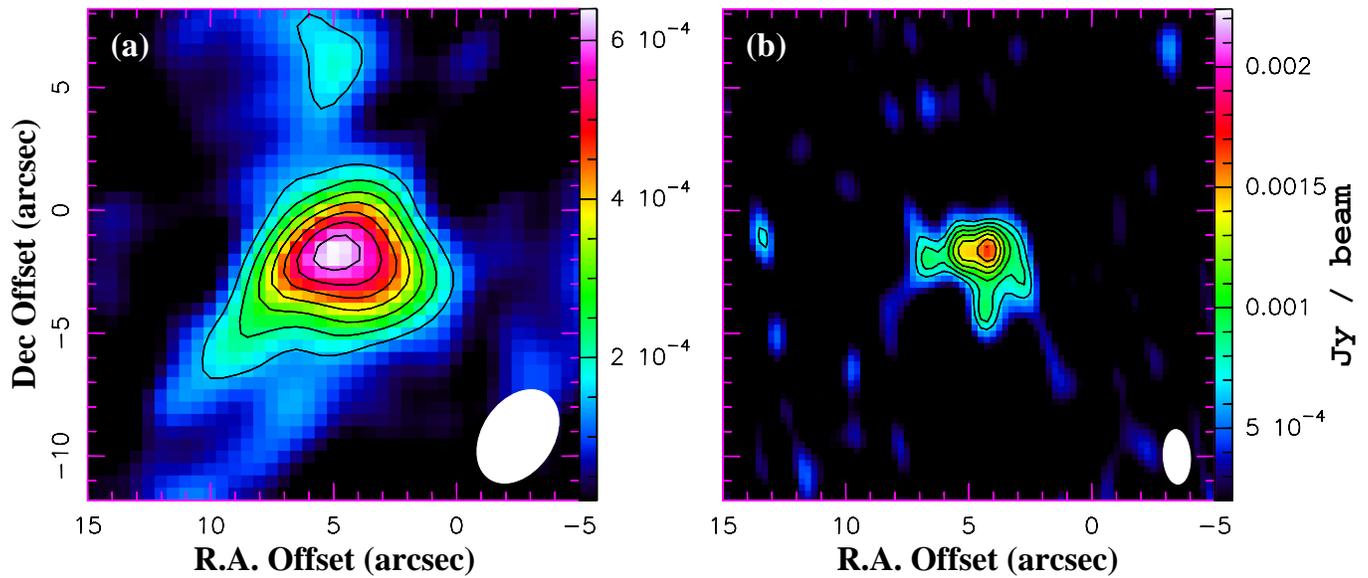}}
\caption{Continuum emission from the Boomerang: (a) 2.6\,mm (contour levels go from $0.15\times10^{-3}$\,Jy\,beam$^{-1}$ to
$0.675\times10^{-3}$\,Jy\,beam$^{-1}$ in steps of $0.075\times10^{-3}$\,Jy\,beam$^{-1}$) and (b) 1.3\,mm (contour levels go from
$0.666\times10^{-3}$\,Jy\,beam$^{-1}$ to $1.67\times10^{-3}$\,Jy\,beam$^{-1}$ in steps of $0.167\times10^{-3}$\,Jy\,beam$^{-1}$). The restoring beams
are shown in the bottom right corner of each panel. 
}
\label{cont}
\end{figure}

The continuum image at 1.3 (2.6) mm, with a beam of $2\farcs2\times1\farcs1$ ($4\farcs1\times2\farcs9$) having its major-axis at $PA=-176.6\arcdeg$
($-33.9\arcdeg$), shows a compact source, largely unresolved, with evidence for some extended weak 
emission (Fig.\,\ref{cont}). It is interesting that the 1.3 mm image shows a narrow linear feature extending about $3{''}$ south from the central
peak, that, although of marginal significance, is aligned with the knotty, collimated feature seen in the HST image. The peak fluxes extracted from
the 1.3 and 2.6\,mm continuum images, with the former convolved to the same beam as the latter, are 3.64 mJy and 0.64 mJy, respectively.

The peak of the continuum emission, Pk$_{mm}$, as measured in the 1.3 and 2.6\,mm continuum images, has J2000 coordinates of R.A.$=12^h44^m45.99^s$,
Dec$=-54^\circ31'13.5{''}$, with an error of about $\pm0\farcs15$. This position is significantly offset from the phase-center. It is near to, but
slightly offset to the west (by about $0\farcs9$) from, the location of the central point source in the Boomerang seen at near-IR to mid-IR
wavelengths (Pk$_{ir}$), that has 2MASS coordinates of R.A.$=12^h44^m46.09^s$, Dec$=-54^\circ31'13.3{''}$ (the
DENIS, WISE and Akari-Mid positions of Pk$_{ir}$ are all within $\sim0\farcs15$ of the 2MASS one). Pk$_{mm}$ is also offset westwards from the
long-axis of the nebula seen in CO\,(J=2-1) by about $1{''}$. The brightness distribution of CO\,(J=2-1) in the waist 
is also not symmetric about the nebular axis, and its peak, at (and near) the systemic velocity is located at the same position as Pk$_{mm}$ (within
measurement errors).

Our ATCA data show no detectable emission at the position of the Boomerang Nebula with $1\sigma$ upper limits of 
$23, 34,$~and $21~\mu$Jy/beam at 5.5, 9, and 23\,GHz, respectively.


\section{Discussion}\label{discus}
The ALMA data validates some basic features of the SN97 model, which consists of two nested spherically symmetric
shells: a warm inner shell extending 2.5\arcsec--6\arcsec~with an expansion
velocity of about 35\,\kms, and a cool, extended outer shell extending 6\arcsec--33\arcsec, with a
velocity of about 164\,\kms. The latter shell is cooled below the temperature
of the CMB through adiabatic expansion. The ALMA observations show that the inner
component is bipolar, with a dense waist, and the outer component is
patchy, but roughly circular and similar in dimensions to the model,
bearing in mind that a significant fraction of the flux in absorption
has been resolved out. 

Although SN97 adopted a constant expansion velocity for the ultra-cold outflow in their model, they admitted the possibility that the outflow velocity
increased gradually from 35\,\kms~in the inner shell to 164\,\kms~in the outer one. Our ALMA data reveal a radially-varying outflow velocity in the
ultra-cold outflow, suggesting that the absorbing material in this outflow was either
ejected with a velocity that has varied with time or in a single, ``explosive" event with a wide distribution of velocities. In either case, the
adiabatic cooling due to expansion will still occur, but the temperature profile of SN97 with a constant velocity may need to be modified.

The good correspondence between the limb-brightened bipolar lobes seen in the ALMA CO\,(J=2-1) and HST images confirms that these are hollow
cavities with dense walls of molecular gas and dust producing both the molecular-line structures seen at millimeter wavelengths and the scattered-light structures 
seen at visible wavelengths. The molecular gas associated with the diffuse visible-light nebulosity surrounding the N- and S-lobes is not seen in the 
CO\,(J=2-1) map, presumably because it belongs to the ultra-cold outflow and does not emit in the J=2-1 line. 

The diffuse biconical morphology seen in the visible is likely 
the result of preferential illumination of the ultra-cold outflow (that is intrinsically spherical) due to the presence of a central dusty structure 
detected in the continuum images, which hinders the starlight from escaping at low-latitudes (the long axis of the nebula defines the polar axis).
Several radial streaks seen in the diffuse nebulosity provide support for this scenario, as these may then be simply explained as
``shadows" due to inhomogeneous dust obscuration present around the central star. Fainter, smooth nebulosity is seen at low-latitudes, and
is likely due to illumination of circumstellar material there by multiply-scattered starlight.

The central, dusty structure that results in the preferential illumination of the ultra-cold outflow, may be a flared disk. Alternatively, since the
opening angles of the diffuse nebulosity and the V-shaped bases of the lobes, seen in the visible, appear to be similar, it is also plausible that 
the central dusty region was formed by heavy spherical mass-loss, but has a biconical physical cavity in it through which the starlight streams out.
Such a cavity may have been carved out by a bipolar jet -- e.g., for the pre-planetary nebula Hen\,3-401, which also has cylindrical lobes with
V-shaped bases, Sahai, Bujarrabal, \& Zijlstra (1999) conclude that the lobe shapes may be produced by a fast, momentum-conserving jet sweeping-up
material in an ambient medium with a radially decreasing density. 

We suggest that a similar mechanism may have produced the central bipolar nebula in the Boomerang as well. This mechanism would require the jet
velocity, $V_{jet}$, to be larger than the expansion velocity, $V_{amb}$, of the ambient circumstellar material (i.e., the ultra-cold outflow in the
case of the Boomerang), and the expansion velocity, $V_{lob}$, in the swept-up shell (i.e., the lobes) to be  $<V_{jet}$ and $>V_{amb}$. This would
have been a problem if the ultra-cold outflow had a radially-constant outflow velocity of 
164\,\kms~since the expansion velocity in the lobes is much lower. Thus our finding that the ultra-cold outflow is likely expanding at relatively low
speeds in its inner regions is crucial for supporting the jet-interaction scenario that we have proposed for producing the central bipolar nebula in
the Boomerang. The linear, knotty feature seen in the HST image (Fig.\,\ref{hst606w}, inset) may arise due to line emission (e.g., H$\alpha$) from the
inner regions of this jet. But in order to test this idea, we will require spatially-resolved spectroscopy of this feature, and/or determination of
proper motion in the knots showing high velocities, using HST.


\subsection{Re-Heating of the Ultra-cold Outflow: Photoelectric Dust Grain Heating}
Inspection of the patchy regions of weak emission in the outer regions of the ultra-cold shell in the ALMA CO\,(J=1-0) map and in the ALMA+SEST map
shows
that these patches, although brighter than their surroundings, remain in absorption against the microwave background.
The reason they appear in emission in the ALMA data is because a significant part of the ultracool-outflow absorption structure is smooth
and resolved out by the interferometer. The presence of these regions can be a result 
of local variations (decreases) in column densities (thus producing less
absorption) or warmer temperatures. However, the specific spatial distribution of these 
(i.e., generally at larger radii than the absorption 
patches) suggests that they are due to warmer temperatures because the ultra-cold outflow is
expected to ultimately begin heating up due to heating via
grain photoelectric heating, followed by heating due to photodissociation of CO (e.g., Huggins, Olofsson \&, Johansson 1988). The latter is not
expected to be 
a significant contributor at radii smaller than the photodissociation radius of the self-shielded CO in the outer shell, 
about $5\times10^{18}$\,cm (Knapp \& Chang 1985).

\begin{figure}[!ht]
\resizebox{0.8\textwidth}{!}{\includegraphics{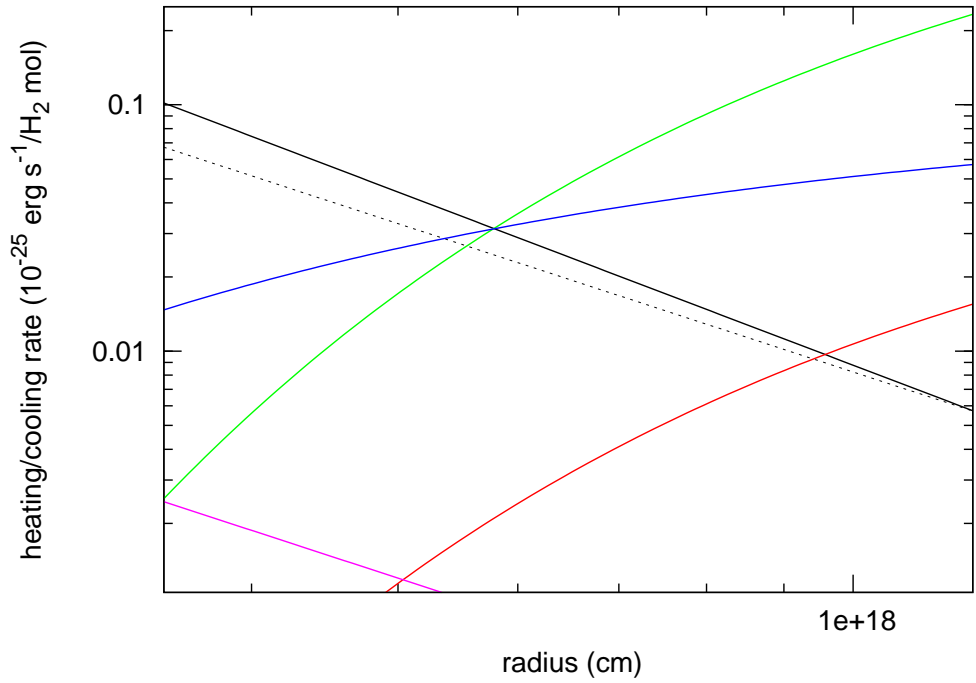}}
\caption{The adiabatic cooling rate ($Q_{adiab}$, black curves), compared to various heating rates per $H_2$ molecule, for the ultracold outflow
in the Boomerang. The solid (dashed) black curve shows the adiabatic cooling rate for a constant (radially-increasing) outflow velocity (see
text in \S\,4 for details). The red and green curves show $Q_{pe}$ for a standard value of the unshielded photoelectric rate, 
$k_{pe}=10^{-26}$\,ergs\,s$^{-1}$,
and a much larger value, $1.5\times10^{-25}$\,ergs\,s$^{-1}$, respectively, and the magenta curve shows the dust-gas frictional heating rate; for each
of these, the dust-to-gas ratio is 1/200. The blue curve shows $Q_{pe}$ for a dust-to-gas ratio lower by a factor 3.3, and the standard $k_{pe}$.
}
\label{peheat}
\end{figure}

We can roughly estimate the radius, $r_{pe}$, at which grain photolelectric heating starts becoming effective, by 
equating the adiabatic cooling rate per $H_2$ molecule, $Q_{adiab}=2\,k_B\,V_{exp}\,T_{kin}(r)/r$, with 
the photoelectric heating rate, $Q_{pe}=k_{pe}\,exp[-\tau$(1000\AA)], where $k_B$ is the Boltzmann constant, $V_{exp}$ is the outflow velocity, 
$T_{kin}$ is the kinetic temperature, $k_{pe}$ is the unshielded heating rate (in ergs\,s$^{-1}$), and $\tau$(1000\AA) is the circumstellar dust
optical depth to the interstellar UV radiation field at a nominal wavelength of 1000\,\AA.  
Substituting $T_{kin}(r)=2.8\,K(-r/1.35\times10^{17})^{-4/3}$ (SN97), 
and $\tau$(1000\AA)$=3.44\times10^{21}\,\mdot(100/\delta)/r$ (Morris \& Jura 1983), with the gas-to-dust ratio, $\delta=200$, and
$\mdot=1.3\times10^{-3}\,\my$, we compare $Q_{pe}(r)$ to $Q_{adiab}$, for 
$k_{pe}=10^{-26}$ (e.g., Huggins et al. 1988), and find that $r_{pe}\sim10^{18}$\,cm (Fig.\,\ref{peheat}), larger than the observed value 
($\sim25{''}$, or $0.56\times10^{18}$\,cm) by almost a factor 2. This could be accommodated, e.g., by an increase in $k_{pe}$ by a factor $\sim$15, or
an increase in the gas-to-dust ratio by a factor of $\sim$3 (Fig.\,\ref{peheat}). Given the uncertainties in the envelope
parameters and the dust grain properties of such a high velocity
outflow (e.g. very small grains can lead to values of $k_{pe}$ as high as $10^{-25}$, Jura 1976), external
photo-electric heating is clearly a viable mechanism for the outer
emission. A detailed calculation would also need to take into account
the clumpy structure observed in the gas.

The above discussion is based on a constant expansion velocity, but the CO (J=1-0) observations (\S\,\ref{coj10}) suggest that there is a range of
velocities present in the outflows. In order to investigate how this might affect the cooling we consider a power law, V
However, since the CO\,(J=1-0) spectrum suggests that the expansion velocity of the slow outflow is not constant, we assume that it varies as a
power-law, 
$V'(r)=V_o\,(r/r_o)^{\alpha}\,\kms$, and re-derive the radial dependence of the kinetic temperature and adiabatic cooling analytically. 
With $Q'_{adiab}=2\,k_B\,V(r)\,T_{kin}(r)\,(1\,+\,1/2\,d\,lnV/d\,lnr)/r$ (e.g., Goldreich \& Scoville 1976), and assumed to dominate the
heating-cooling terms\footnote{quite reasonable, since SN97 showed that the dust frictional heating terms is much smaller, by a factor
$\sim50$ at $r=10^{17}$\,cm, in the constant outflow velocity model}, we find that $T'_{kin}(r)=T_i\,(r/r_i)^{-(4+2a)/3}$ (i.e., the kinetic
temperature falls more steeply with radius than in the constant expansion velocity case). A determination of $T_0$ will require a new
radiative-transfer/thermodynamic model to fit the SEST single-dish data (as in SN97), that is outside the scope of this paper (and deferred to after
we obtain Cycle 1 data). However, since the average kinetic temperature in the ultra-cold outflow is constrained by the observed absorption, we
estimate $T_0$ by making the simple assumption that 
$T'_{kin}(r)$ at some nominal average radius, $r_{av}$ is the same as that for the constant expansion case. We set 
$V'(r)=164\,\kms\,(r/r_o)^{\alpha}$, with $\alpha=1$, and take $r_{av}$ to be the geometric average of the inner and outer radii of the ultra-cold
outflow in the SN97 model (i.e., $r_i=6{''}$ and $r_o=33{''}$, respectively) , giving $T_0=4.9$\,K. The resulting radial variation of $Q'_{adiab}$,
shown in Fig.\,\ref{peheat}, is not very different from $Q_{adiab}$. 

Finally, we note that since an outflow with a mass-loss rate that is constant with time and a radially-increasing outflow velocity, offers less
optical depth to the interstellar UV radiation field in its outer regions compared to one where the outflow velocity is constant, the discrepancy
that we found above between the observed and estimated values of $r_{pe}$ is likely to  be smaller.

\subsection{Thermal Dust Emission from the Waist}
\begin{figure}[!ht]
\resizebox{0.9\textwidth}{!}{\includegraphics{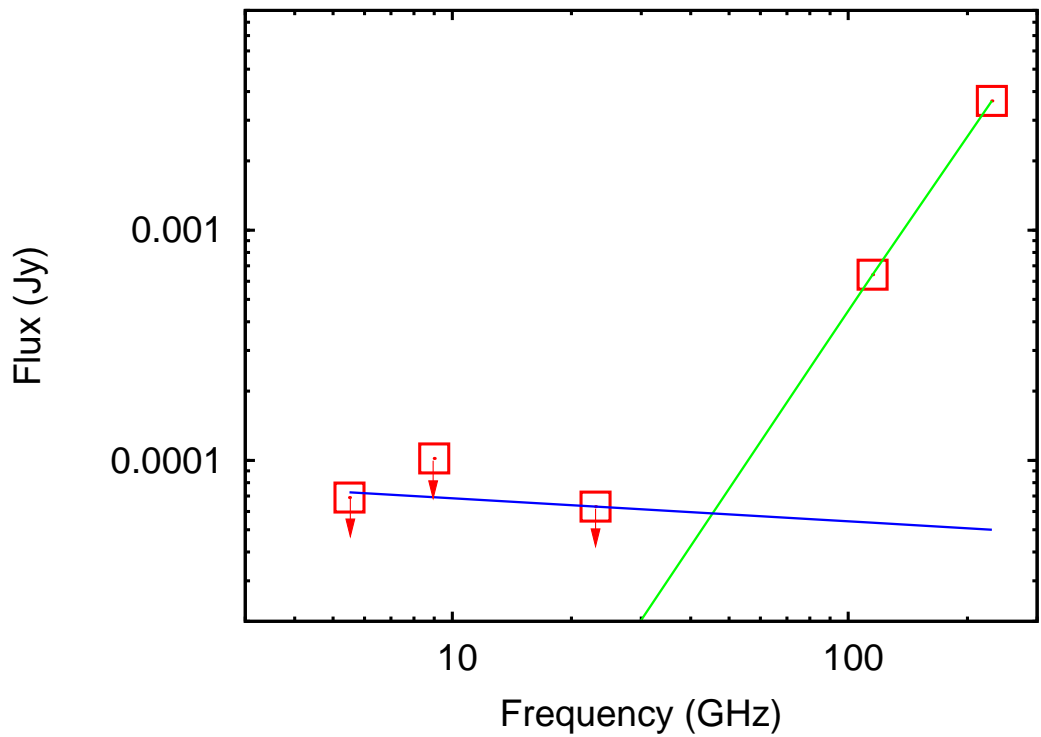}}
\caption{ALMA mm-wave continuum fluxes for the central source and 3-sigma upper limits from ATCA. The data (red squares) have been fitted
with a 2-component model consisting of optically-thin (1) thermal dust emission from mm-sized grains (thus with a shallow emissivity
power-law index 0.6) at a temperature of 45 K (green curve), and (2) free-free emission (blue curve).
}
\label{sed}
\end{figure}

We fit the ratio of the peak 1.3 and 2.6\,mm continuum fluxes, measured within the same  
beam of size $4\farcs1\times2\farcs9$ (see \S\,\ref{contxt}), $R=5.7$ (although S/N is modest and 1.3 mm
UV-coverage is still sparse) as follows. Assuming a power-law dust emissivity, i.e. $\kappa(\nu)\propto\nu^\beta$, we find
that $\beta=0.5$ in the Rayleigh-Jeans (R-J) limit, since $R(\lambda _1/\lambda _2) = (\lambda _1/\lambda _2)^{(2+\beta)}$. If we drop the R-J
approximation, then determining $\beta$ requires a knowledge of the dust temperature, $T_d$. Taking  
$\beta$ = 0.6, 1, 1.5, we get $T_d=45$\,K, 9.5\,K, 5.0\,K and $r_d=1\farcs9, 236{''},
6800{''}$, respectively, using $r_d=(L_* T_*^\beta/16\,\pi\,\sigma)^{1/2}\,T_d^{-(2+\beta/2)}$, where $L_*$ and $T_*$ are the stellar luminosity and
temperature (e.g., Herman, Burger \& Penninx 1986). Since the continuum emission arises from a compact source
with an observed value of $r_d\sim2{''}$, we conclude $\beta\sim0.6$ and $T_d\sim45$\,K (see Fig.\,\ref{sed}). 
Realistically, extinction and reddening of starlight needs to be considered, allowing
somewhat higher $\beta$ and lower $T_d$ values; 
e.g., if only 10\% of the total stellar flux (reddened to 900\,K) is available for heating the dust, then $\beta\sim0.7$ and $T_d\sim23$\,K.
The low value of $\beta$ suggests that the grains in the waist producing the observed emission are 
several mm in size (e.g., Draine 2006). The radio continuum
observations at 5.5, 9, and 23\,GHz show no source, so free-­free emission is unlikely to contribute to the 1.3 and 2.6
mm fluxes (Fig.\,\ref{sed}). 

There is now a growing body of observational evidence for such large grains in the dusty equatorial regions of 
post-AGB objects (e.g., see Sahai et al. 2011) that has opened up an exciting new 
opportunity to probe the very early stages of
grain coagulation, i.e. on very short time-scales (i.e., $\sim 1000$ yr), which is not possible with studies of planet-forming
disks, since the latter are typically $\gtrsim10^6$\,yr old (Sahai et al. 2009).

The mass of dust in the central source derived from the above fluxes and taking $T_d=45$\,K, is $M_d\sim3.5\times10^{-4}$\,\ms, and the total mass is
$M\sim0.071$\,\ms, assuming a dust opacity of $\kappa(1.3mm)\sim1.5$\,cm$^2$\,g$^{-1}$ and the gas-to-dust ratio, $\delta=200$. The expansion time
scale for the
dust-continuum emitting region, assuming its size to be the geometric average of the major and minor axis of the elliptical beam used to measure the
dust continuum ($3\farcs45$ or $7.76\times10^{16}$\,cm), and an expansion velocity equal to half the FWZI of the CO\,(J=2-1) line profile of the
waist,
32.5\,\kms, is about 400\,yr, implying a mass-loss rate $\sim1.8\times10^{-4}$\,\my. This rate is comparable to that derived by SN97 for the inner
shell from the single-dish CO data, $\gtrsim10^{-4}$\,\my. 


\section{Conclusions}\label{conclude}
We have obtained ALMA maps of millmeter-wave line and continuum emission at the highest angular resolution to-date of the Boomerang Nebula, 
the coldest known object in the Universe. We compare these to high-resolution imaging at 0.6\micron~obtained with HST. Our principal
conclusions are:\\
\begin{enumerate}
\item The Boomerang Nebula consists of a central, expanding hourglass-shaped lobe structure with a dense waist, that is seen in emission in CO\
(J=2-1),
surrounded by a patchy, ultra-cold outflow seen in absorption against the microwave backgroud in the CO\,(J=1-0) line. 
\item The HST imaging reveals the precise geometrical shapes of the lobes, which flare out from the center in a
wide-V shape, and then follow a roughly cylindrical shape. Each limb-brightened lobe is embedded in a
larger, more diffuse conical nebulosity with an opening angle that is the same as that of its wide V-shaped inner region.
\item There is good correspondence between the limb-brightened hourglass lobes seen in CO\,(J=2-1) and the visible-light imaging, showing that these are
hollow cavities with dense walls of molecular gas and dust. The diffuse biconical nebulosity seen in the visible surrounding these limb-brightened lobes is
likely the result of preferential illumination of the ultra-cold outflow by light from the central star due to the presence of the central dusty
structure, seen in the continuum images.
\item The ratio of the 1.3\,mm to the 2.6\,mm continuum flux, $R=5.7$, is low compared to the expectation from thermal emission by small grains. Radio
observations at 5.5, 9, and 23 GHz show no source, so free-­free emission does not contribute to the mm-wave fluxes. We infer 
a dust temperature of $\sim45$\,K and a power-law dust emissivity exponent $\beta\sim0.6$; the low value of $\beta$ suggests that the grains in the
waist producing the observed emission are several mm in size. 
\item The total mass associated with the dusty waist is $M\sim0.071$\,\ms (assuming a gas-to-dust ratio, $\delta=200$ and a dust opacity of
$1.5$\,cm$^2$\,g$^{-1}$ at 1.3mm). The expansion time scale for
dust-continuum emitting region is about 400 yr, implying a mass-loss rate $\sim1.8\times10^{-4}$\,\my.
\item A comparison of the single-dish and ALMA CO\,(J=1-0) map shows that a large fraction of the absorption signal from the ultra-cold outflow is 
resolved out in our ALMA CO\,(J=1-0) map. The observed absorption
regions are seen to extend over a region $\sim50{''}$ in size, beyond which can be seen faint patchy ``emission" regions, lying within a region of
size  $\sim65{''}$.  
\item The combined ALMA+SEST map shows that the outer patchy ``emission" regions, although brighter than their surroundings, remain in absorption
against the microwave background; the presence of these regions is consistent with the re-warming of the ultra-cold gas due to photoelectric grain
heating.
\item The CO\,(J=1-0) spectra towards the center of the nebula, extracted from the ALMA+SEST map, shows absorption over a wide
range of velocities, implying that the absorbing material in the ultra-cold outflow was
either ejected with a velocity that has varied with time, or in a single, explosive event with a wide distribution of velocities.
\item The central bipolar nebula is likely the result of a fast, momentum-conserving bipolar jet interacting with the relatively slowly expanding
material present in the inner regions of the ultra-cold outflow. This jet may also be responsible for carving out biconical cavities in a central
dusty structure through which the starlight streams out preferentially to produce the large-scale diffuse biconical appearance of the Boomerang
Nebula. Plausible evidence for such a jet is provided by the presence of a linear, knotty feature emanating from the central star, seen in the HST
image.
\end{enumerate}

\section{Acknowledgments}  This paper makes use of the following ALMA data: 
ADS/JAO.ALMA\#2011.0.00510.S. ALMA is a partnership of ESO (representing its 
member states), NSF (USA) and NINS (Japan), together with NRC (Canada) and 
NSC and ASIAA (Taiwan), in cooperation with the Republic of Chile. The Joint 
ALMA Observatory is operated by ESO, AUI/NRAO and NAOJ. The National Radio Astronomy Observatory is a facility of the National Science Foundation
operated under cooperative agreement by Associated Universities, Inc. RS's contribution to the research described here
was carried out at JPL, California Institute of Technology, under a contract with NASA and
partially funded through the internal Research and Technology Development program. WV acknowledges support by
the Deutsche Forschungsgemeinschaft (DFG; through the Emmy Noether
Research grant VL 61/3-1) and Marie Curie Career Integration Grant 321691. This work was supported in part by NSF grant AST 08-06910 (to PJH).

\end{document}